\documentclass[fleqn,usenatbib]{mnras}

\usepackage{newtxtext,newtxmath}

\usepackage[T1]{fontenc}

\DeclareRobustCommand{\VAN}[3]{#2}
\let\VANthebibliography\thebibliography
\def\thebibliography{\DeclareRobustCommand{\VAN}[3]{##3}\VANthebibliography}

\usepackage{graphicx}	
\usepackage{amsmath}	
\usepackage{siunitx}
\usepackage{siunitx}
\usepackage{orcidlink}

\title[Rapid late-time reionization]{\boldmath Rapid late-time reionization: constraints and cosmological implications}

\author[W. Elbers]{Willem Elbers\orcidlink{0000-0002-2207-6108}\thanks{E-mail:willem.h.elbers@durham.ac.uk}
\\
Institute for Computational Cosmology, Department of Physics, Durham University, South Road, Durham, DH1 3LE, UK
}

\date{Accepted XXX. Received YYY; in original form ZZZ}

\pubyear{\the\year{}}

\begin{document}
\label{firstpage}
\pagerange{\pageref{firstpage}--\pageref{lastpage}}
\maketitle

\begin{abstract}    
We present constraints on the reionization optical depth, $\tau$, obtained using several independent methods. First, we perform a non-parametric reconstruction of the reionization history, using Lyman-$\alpha$ constraints on the evolution of the volume-averaged neutral hydrogen fraction, $x_\mathrm{HI}(z)$, including recent results from \emph{JWST}. When combined with baryon acoustic oscillation (BAO) measurements from DESI and Big Bang nucleosynthesis constraints, these data imply a rapid reionization history ($z_\mathrm{mid}=6.99^{+0.12}_{-0.18}$ and $\Delta z_{50}=1.09^{+0.10}_{-0.28}$) and a value of $\tau=0.0491^{+0.0014}_{-0.0031}$, which is largely insensitive to the assumed cosmological model and independent of cosmic microwave background (CMB) data. The optical depth can also be measured from large-scale $(\ell<30)$ CMB polarization data, yielding constraints that are similarly model-insensitive and consistent with the Ly$\alpha$ bound. Third, $\tau$ may be constrained from the attenuation of small-scale $(\ell>30)$ CMB anisotropies, but the results are sensitive to the choice of cosmological model. Assuming $\Lambda$CDM and combining small-scale CMB data with CMB lensing and type 1a supernovae (SNe) yields tight constraints that are compatible with the Ly$\alpha$ bound. Adding galaxy clustering and lensing measurements brings the constraints further into agreement with the Ly$\alpha$ bound. These independent results reinforce a consensus picture in which reionization is rapid and late. However, the combination of small-scale CMB, CMB lensing, and BAO data yields $\tau=0.094\pm0.011$, which is in $4\sigma$ tension with our Ly$\alpha$ bound. Non-standard reionization scenarios, or relaxing our assumption that the onset of reionization occurs after $z=15$, can reconcile some but not all constraints. Concordance is restored in alternative cosmological models, such as models with dynamical dark energy favoured by BAO, CMB, and SNe data.
\end{abstract}




\section{Introduction}\label{sec:intro}

The epoch of reionization was a cosmic phase transition in which radiation from the first luminous objects ionized the neutral intergalactic medium (IGM) \citep{Loeb01,Robertson10,Choudhury22}. Establishing the timeline of reionization is a central challenge in cosmology, with profound implications for our understanding of the first stars and our interpretation of the cosmic microwave background (CMB). This primordial signal was perturbed at late times due to Thomson scattering by free electrons generated during reionization. The amount of scattering, characterized by the optical depth, $\tau$, is degenerate with other cosmological parameters of interest. Hence, reliable constraints on reionization are indispensable for cosmological model inference.

In recent years, a wide array of observations have made it possible to piece together the global history of reionization \citep[e.g.][]{Greig17,Dai19,Mason19b,Hazra20,Paoletti25,Sims25}. Information on the timing of reionization can be derived from many different observables, such as damping wings in quasar and galaxy spectra \citep[e.g.][]{Mesinger07,Davies18,Greig22,Umeda24,Zhu24}, equivalent widths of Lyman-$\alpha$ emission lines \citep[e.g.][]{Mason18,Hoag19,Bruton23,Tang24}, the fraction of dark pixels in the Lyman-$\alpha$ and Lyman-$\beta$ forests \citep[e.g.][]{McGreer15,Jin23}, the clustering of Lyman-$\alpha$ emitters \citep[e.g.][]{Ouchi10,Sobacchi15}, the kinematic Sunyaev-Zeldovich (kSZ) effect \citep[e.g.][]{Zahn12,Smith17,SPT24_Reion}, the post-reionization Lyman-$\alpha$ forest \citep[e.g.][]{Cen09}, and the 21-cm power spectrum \citep[e.g.][]{Abdurashidova22,Ghara25}. The arrival of the \emph{James Webb Space Telescope} (\emph{JWST}) has enabled observations of the onset of reionization at $z>10$ \citep[e.g.][]{Bruton23,CurtisLake23,Hsiao24,Munoz24,Tang24,Cohon25,Llerena25,Umeda25}.

The strongest CMB constraints on $\tau$ derive from large-scale polarization data. Thomson scattering of the CMB quadrupole results in linear polarization on the horizon scale, which produces a characteristic feature in the polarization power spectrum on large scales \citep{Hu97,Reichardt16}. Detection of this reionization feature was first achieved with the \emph{Wilkinson Microwave Anisotropy Probe} (WMAP) \citep{Kogut03} and measurements were further refined with \emph{Planck} \citep{Planck14,Planck16_Reionization,Planck20}, culminating in a precise determination of $\tau=0.051 \pm 0.006$ \citep{Planck20_tau}. Recently, the polarization signature was detected with the CLASS telescope \citep{Li25}, yielding $\tau=0.053_{-0.019}^{+0.018}$, in agreement with the \emph{Planck} value, though with larger errors. On small scales, the first-order effect of reionization on the CMB is an overall attenuation of the signal, introducing a degeneracy between $\tau$ and the amplitude of the primordial power spectrum, $A_\mathrm{s}$. When small-scale CMB data are complemented with low-redshift probes to break this degeneracy, the resulting constraints on $\tau$ are competitive with other approaches \citep{Giare24b}.

Among low-redshift probes that can be combined with CMB data, measurements of baryon acoustic oscillations (BAO) are the most powerful \citep{Eisenstein98}. They provide a robust geometric probe of the expansion history that effectively breaks parameter degeneracies. However, the latest BAO measurements by the Dark Energy Spectroscopic Instrument \citep{DESI24.KP7B,DESI24.KP7A,DESI25.I,DESI25.II} are in tension with CMB experiments \citep{ACT_DR6_Spectra25,GarciaQuintero25,Camphuis25}. This tension is at least partially responsible for anomalies, such as the preference for excess gravitational lensing \citep{Calabrese08,Planck20,Mokeddem23} and the tension between cosmological constraints on the sum of neutrino masses and bounds from neutrino oscillation experiments \citep{Craig24,Loverde24,Elbers25,Elbers_DESI25,Green25}. Notably, it contributes to the $3\sigma$ preference for dynamical dark energy from the combination of DESI BAO and CMB data \citep{DESI25.II}. To make progress on fundamental questions about neutrinos and the nature of dark energy, it is crucial to understand the origin of these anomalies. Recently, \citet{Sailer25} and \citet{Jhaveri25} pointed out that the tension between DESI and CMB data could be explained by unknown systematics in the large-scale polarization measurements of $\tau$, potentially allowing a greater optical depth of $\tau=0.09$. This value is much larger than predicted by standard reionization models, but could reconcile CMB measurements of the expansion history with those from DESI.

At the same time, new \emph{JWST} observations suggest that galaxies were more efficient at producing ionizing radiation during the epoch of reionization \citep{Atek24,Simmonds24,Llerena25,Pahl25,Papovich25}. How much of this radiation escapes into the IGM is crucial for reionization, but cannot be measured directly at high redshifts. The application of scaling relations, calibrated to low-redshift galaxy samples, suggests that the escape fraction was greater for reionization-era galaxies and increasing towards the faint end of the luminosity function \citep{Chisholm22,Mascia24}. As shown by \citet{Munoz24}, the increased ionizing efficiencies and escape fractions, combined with incomplete observations at the faint end, could lead to an overabundance of ionizing photons, an early end to reionization, and a greater optical depth, $\tau=0.09$, in tension with CMB and Lyman-$\alpha$ observations (termed a `photon budget crisis'). However, more recent \emph{JWST} studies have inferred the escape fractions of reionization-era galaxies by fitting their spectral energy distributions and found no relation between the escape fraction and UV brightness \citep{Papovich25,Giovinazzo25}, calling into question the use of relations derived from low-redshift samples. There is still considerable uncertainty regarding the escape fraction, but both studies agree that the average value is sufficiently low for a late end to reionization around $z\approx5-6$, apparently resolving the photon budget crisis (see also \citealt{Pahl25}).

Motivated by anomalies in the latest cosmological data and the new \emph{JWST} observations, this paper presents constraints on $\tau$ employing several independent methods to minimize the dependence on any single dataset. First, we perform a non-parametric reconstruction of the reionization history using astrophysical constraints on the evolution of the volume-averaged neutral hydrogen fraction, $x_{\mathrm{HI}}(z)$, including the latest \emph{JWST} measurements. A novelty of our analysis is that we simultaneously vary the reionization history and cosmological parameters, without relying on CMB data. The optical depth, $\tau$, is given by
\begin{align}
    \tau = n_\mathrm{H}c\sigma_\mathrm{T}\int_0^{z_\mathrm{max}}\mathrm{d}zx_\mathrm{e}(z)\frac{(1+z)^2}{H(z)}, \label{eq:tau}
\end{align}

\noindent
where $n_\mathrm{H}$ is the present-day density of hydrogen, $\sigma_\mathrm{T}$ the Thomson scattering cross section, $x_\mathrm{e}(z)$ the free electron fraction, and $H(z)$ the Hubble rate. We can determine $\tau$ independently of the CMB by combining constraints on $x_\mathrm{HI}(z)$ with BAO and Big Bang nucleosynthesis (BBN) measurements. Together, these data effectively constrain the hydrogen density, $n_\mathrm{H}$, and the expansion history, $H(z)$. We will compare the results of this reconstruction with constraints on $\tau$ obtained from small-scale and large-scale CMB data in combination with various low-redshift probes.

The remainder of the paper is structured as follows. We describe the data used in our analysis in Section~\ref{sec:data}. The non-parametric reconstruction of the reionization history is presented in Section~\ref{sec:nonpar}. In Section~\ref{sec:cmb}, we compare the results with constraints on $\tau$ from the CMB and present results for extended cosmological models. Finally, we present our conclusions in Section~\ref{sec:discussion}.

\section{Data}\label{sec:data}

We make use of both astrophysical constraints on reionization and various other cosmological datasets as outlined in the two subsections below. These data are used for Bayesian cosmological parameter inference, employing a Markov Chain Monte Carlo algorithm implemented in the \texttt{cobaya} code \citep{Torrado11}. Cosmological calculations are performed with a modified version of the \texttt{CLASS} code \citep{Lesgourgues11}. The chains are finally analyzed with the \texttt{getdist} package \citep{Lewis19}.

\subsection{Reionization data}

The primary dataset used in our analysis is a compilation of recent constraints on the evolution of the volume-averaged neutral hydrogen fraction, $x_{\mathrm{HI}}(z)$, from the literature. The constraints are listed in Table~\ref{tab:data} and derive mostly from damping wing features in quasar and galaxy spectra and equivalent widths of Lyman-$\alpha$ emission lines. We also use the Lyman-$\alpha$ and Lyman-$\beta$ forest dark pixel fraction constraints of \citet{Jin23}, which are similar to but more conservative than the commonly-used constraints from \citet{McGreer15}. In compiling this dataset, we gave priority to data points that were more recent. We also required that the probability density was publicly available (with one exception discussed below) and that the data were independent. However, it has been pointed out by \citet{Kageura26} that the spectrum used in \citet{CurtisLake23} was also used by \citet{Umeda25}. We show in Section~\ref{sec:baseline} that excluding the constraint from \citet{CurtisLake23} does not significantly impact our results.

The values in Table~\ref{tab:data} are given only for reference. In practice, we sample directly from the marginalized posterior probability density, $p(x_{\mathrm{HI}})$. Redshift errors are modelled by including one nuisance parameter for each data point, imposing a uniform prior, $\Delta z\in[-z_\text{min}, z_\text{max}]$, with the values given in Table~\ref{tab:data}. This is a conservative choice, but only has a small impact on the results. Additionally, the following comments apply to specific data points:

\begin{enumerate}
    \item The probability densities were not available for \citet{Jin23}. Following \citet{Greig17} and \citet{Mason19b}, we model the upper bounds $x_\mathrm{HI}<\mu\pm\sigma$ as one-sided Gaussians with $p(x_\mathrm{HI})=1$ for $x_\mathrm{HI}<\mu$ and $p(x_\mathrm{HI})=\phi_\mathrm{Normal}(x_\mathrm{HI};\mu,\sigma)$ for $x_\mathrm{HI}\geq\mu$. We tested an alternative approach in which the upper bounds were modelled as latent variables, but the results were similar.
    \item The measurements of \citet{Mason18,Mason19} and \citet{Hoag19} assume that the IGM is fully ionized at $z=6$. Following \citet{Mason19b}, we interpret their posteriors, $p(x_\mathrm{HI})$, as posteriors for the difference $\Delta x_{\mathrm{HI}}=x_\mathrm{HI}(z)-x_\mathrm{HI}(6)$.
    \item Following \citet{Greig22}, we average the posteriors from \citet{Greig22} with those from \citet{Yang20} for J1007+2115, \citet{Wang20} for J0252-0503, and \citet{Davies18} for J1120+0641 and J1342+0928 to obtain the $z=7.29$ data point. 
    \item For comparison, we also include the equivalent width constraints from \citet{Tang24} as an alternative at $z>6.5$. The posteriors were shared through private communication.
\end{enumerate}

\begin{table}
    {
    \centering
    \caption{An overview of the astrophysical constraints on the neutral hydrogen fraction used in this paper. For comparison, we also include the constraints from \citet{Tang24}, which are not part of the main sample.}
    \label{tab:data}
    \begin{tabular}{p{3cm}cc}
        \hline
        Reference & $z$ & $x_{\mathrm{HI}}(z)$ \\
        \hline
        \citet{Jin23} & $5.5\pm0.2$ & $<0.09\pm0.08$ \\
        \citet{Jin23} & $5.7\pm0.2$ & $<0.16\pm0.14$ \\
        \citet{Umeda25} & $5.8_{-0.4}^{+0.5}$ & $0.25_{-0.20}^{+0.10}$ \\
        \citet{Jin23} & $5.9\pm0.2$ & $<0.28\pm0.08$ \\
        \citet{Greig24} & $5.90\pm0.05$ & $<0.21$ \\
        \citet{Greig24} & $5.95\pm0.05$ & $<0.20$ \\
        \citet{Greig24} & $6.05\pm0.05$ & $<0.21$ \\   
        \citet{Greig24} & $6.15\pm0.05$ & $0.20_{-0.12}^{+0.14}$ \\
        \citet{Greig24} & $6.35\pm0.05$ & $0.29_{-0.13}^{+0.14}$ \\        
        \citet{Greig24} & $6.55\pm0.05$ & $<0.18$\\        
        \citet{Mason18} & $6.9\pm0.5$ & $0.59_{-0.15}^{+0.11}$ \\
        \citet{Umeda25} & $7.0_{-0.4}^{+0.5}$ & $0.65_{-0.35}^{+0.27}$ \\
        \citet{Greig22} & $7.29\pm0.27$ & $0.49_{-0.14}^{+0.13}$ \\
        \citet{Hoag19} & $7.6\pm0.6$ & $0.88_{-0.10}^{+0.05}$ \\
        \citet{Mason19} & $7.9\pm0.6$ & $>0.76$ \\
        \citet{Umeda25} & $8.6_{-0.9}^{+1.0}$ & $1.00_{-0.20}^{+0.00}$ \\
        \citet{Umeda25} & $10.4_{-0.8}^{+2.0}$ & $1.00_{-0.40}^{+0.00}$ \\
        \citet{Bruton23} & $10.6\pm0.0013$ & $<0.88$\\
        \citet{CurtisLake23} & $11.49_{-0.05}^{+0.04}$ & $0.84_{-0.27}^{+0.14}$\\
        \hline
        \citet{Tang24} & $7.0_{-0.5}^{+1.0}$ & $0.48_{-0.22}^{+0.15}$\\
        \citet{Tang24} & $8.8_{-0.8}^{+1.2}$ & $0.81_{-0.24}^{+0.12}$\\
        \citet{Tang24} & $11.0_{-1.0}^{+2.3}$ & $0.89_{-0.21}^{+0.08}$\\
        \hline
    \end{tabular}
    }
\end{table}

\subsection{Other cosmological data}

Additional data are required to relate the ionization history, $x_{\mathrm{HI}}(z)$, to the optical depth, $\tau$, when varying all cosmological parameters. To do this, we make use of a BBN prior on the physical baryon density, $\Omega_\mathrm{b}h^2$, from \citet{Schoneberg24}.\footnote{When we vary the effective number of relativistic species, $N_\mathrm{eff}$, we also incorporate the covariance between $N_\mathrm{eff}$ and $\Omega_\mathrm{b}h^2$ \citep{Schoneberg24}.} We constrain the cosmic expansion history using the latest BAO data from Data Release 2 (DR2) of the \citet{DESI25.I,DESI25.II}.

We also consider a range of CMB datasets from the \citet{Planck20,Planck20_Spectra}. We specifically use the \texttt{CamSpec} likelihood for the small-scale ($30\leq\ell\leq2500$) auto and cross power spectra of temperature and polarization anisotropies \citep{Efstathiou19,Rosenberg22}, denoted as TTTEEE. When considering large-scale ($\ell<30$) polarization data from \emph{Planck}, denoted as lowE, we compare four different likelihoods to determine the impact of differences in implementation. We consider the \texttt{SimAll} likelihood from the \citet{Planck20_Spectra}, the \texttt{SRoll2} likelihood from \citet{Delouis19}, the \texttt{LoLLiPoP} likelihood from \citet{Tristram21,Tristram24}, and the new \texttt{ELiCA} likelihood from \citet{Genesini26}. We also use small-scale CMB data from the Atacama Cosmology Telescope (ACT) Data Release 6 \citep{ACT_DR6_Spectra25,ACT_DR6_Cosmology25} and the South Pole Telescope (SPT) `SPT-3G D1' \citep{Camphuis25}. The combination of ACT and SPT data is denoted as ACT-SPT, where we neglect their cross correlation due to the small overlap in area \citep{Camphuis25}. Since ACT and SPT do not measure large-scale CMB anisotropies, we also employ a combination of \emph{Planck}, ACT, and SPT denoted as P-ACT-SPT. Specifically, we combine the SPT likelihood with the P-ACT combination that relies on \emph{Planck} on intermediate scales ($30<\ell<600$) and ACT on smaller scales ($\ell\geq600$) \citep{ACT_DR6_Spectra25}.\footnote{The only difference with the P-ACT combination in \citet{ACT_DR6_Spectra25} is that we exclude large-scale temperature data (lowT), which can play a role when lowE is also excluded \citep[e.g.][]{Giare24b}. However, we verified that including lowT only has a small impact once \emph{Planck} or P-ACT is combined with DESI BAO, slightly increasing the tension with our Lyman-$\alpha$ bound.} We also make use of CMB lensing data from \emph{Planck} \citep{Planck20_Lensing,Carron22}, ACT \citep{ACT_DR6_Lensing,Qu24,MacCrann24}, and SPT \citep{SPT_Lensing}, using the extended version of their joint likelihood \citep{ACT_SPT_Planck_Lensing}, denoted simply as Lensing.

As an alternative to BAO data, we also rely on type 1a supernovae (SNe) to constrain the late-time expansion history. In this case, we employ the Dark Energy Survey (DES) Year 5 compilation of 1635 photometrically confirmed SNe combined with 194 low-$z$ external SNe \citep{DES24}. Finally, we make use of a joint analysis of 2-point correlation functions of galaxy positions, galaxy lensing, CMB lensing, and their cross correlations from DES, \emph{Planck}, and SPT \citep{Omori23,Chang23,Abbott23}, which we denote as $6\times2$pt. In the $6\times2$pt likelihood, we adopt the same scale cuts for the dynamical dark energy model $w_0w_a$CDM as for $\Lambda$CDM, following \citet{Abbott23b}. We adapt the public likelihood code to a version compatible with \texttt{cobaya} using the \texttt{CosmoSIS2cobaya} package\footnote{\url{https://github.com/JiangJQ2000/cosmosis2cobaya}}, following \citet{Ye24}.

\section{Non-parametric reconstruction}\label{sec:nonpar}

The reionization history, $x_{\mathrm{HI}}(z)$, has been modelled using a wide range of methods, from simple functional forms to more complex and flexible approaches \citep[e.g.][]{Hazra17,Millea18,Trac18,Sailer22,MonteroCamacho24,Ilic25,Paoletti25}. The most common choice, used by Planck, is to adopt a smooth step function given by
\begin{align}
    x_\mathrm{HI}(z) = \frac{1}{2}\left[1-\tanh\left(\frac{(1+z_\mathrm{mid})^{1.5}-(1+z)^{1.5}}{0.75(1+z_\mathrm{mid})^{0.5}}\right)\right], \label{eq:tanh_form}
\end{align}

\noindent
where $z_\mathrm{mid}$ is the midpoint of reionization. While this approach is simple and useful for many applications, non-parametric methods offer greater flexibility with the distinct advantage of not excluding any reionization histories \emph{a priori} \citep{Mason19b,Krishak21,Cheng25}. We adopt such an approach in this work, but also compare with fits to Eq.~\eqref{eq:tanh_form}.

\begin{figure*}
    \includegraphics[width=\textwidth]{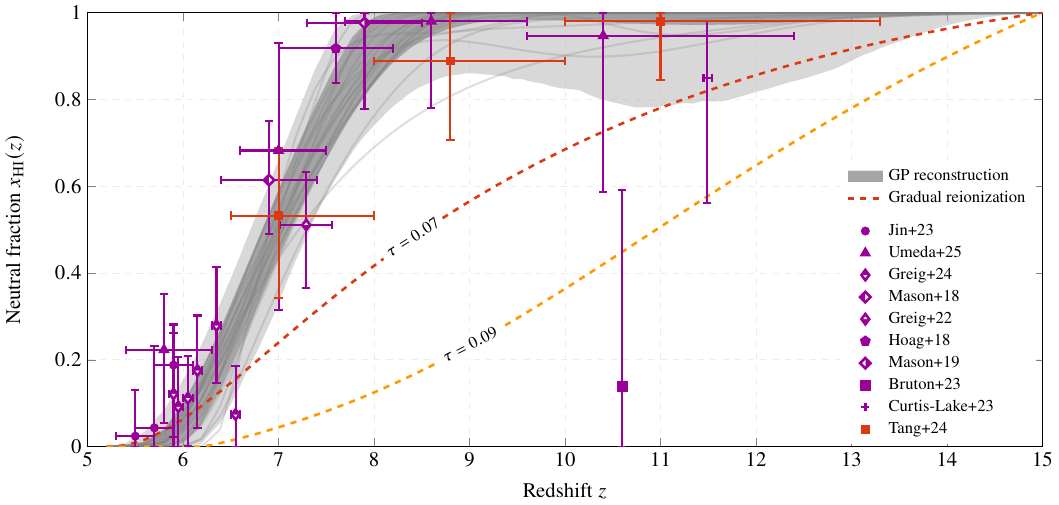}\vspace{-1em}
    \caption{Gaussian process reconstruction of the reionization history using $x_\mathrm{HI}(z)$ constraints. The dark and light regions represent the 68\% and 95\% posterior predictive distributions.
    The faint solid lines are 20 random draws from the distribution.    
    The dashed lines are representative gradual reionization histories that produce large optical depths: $\tau=0.07$ and $\tau=0.09$. These histories are inconsistent with the constraints, particularly around $7\lesssim z\lesssim 8$. 
    The data points indicate the midpoint of the 25\% highest probability density regions (HPD) of the likelihood and the vertical error bars indicate the 68\% HPD regions. The alternative constraints from \citet{Tang24}, shown in red, are not used in the main analysis but are consistent with the main sample.}
    \label{fig:gaussian_reconstruction}
\end{figure*}

\subsection{Methods}

Here, we perform a non-parametric reconstruction of the reionization history using Gaussian process regression. We model the neutral hydrogen fraction by introducing a Gaussian process, $\{\chi_a\}$, where $a=(1+z)^{-1}$ is the scale factor and $X=(\chi_{a_1},\dots,\chi_{a_n})$ is a multivariate Gaussian random variable. The $\chi_a$ are unbounded latent variables that are mapped to the neutral hydrogen fraction, $x_\mathrm{HI}(a)$, by means of a wrapping function $F\colon\mathbb{R}\to(0,1)$.

This method allows us to generate smooth functions by sampling large numbers of points from a multivariate normal distribution with mean, $\mu(a)$, and a covariance matrix, $C(a_i,a_j)$, determined by the kernel function. We use a squared exponential kernel, given by
\begin{align}
    k(a_i,a_j;\sigma,\ell) = \sigma^2\exp\left(-\frac{(a_i-a_j)^2}{2\ell^2}\right),
\end{align}

\noindent
where $\sigma$ is the typical size of fluctuations in $x_\mathrm{HI}$ around the mean. The correlation length is determined by the parameter $\ell$, which we include as a free hyper parameter. Further technical details of the Gaussian process are given in Appendix~\ref{sec:details}.

We impose two boundary conditions at the endpoints of reionization. We require $x_\mathrm{HI}(z_\mathrm{min}) = \epsilon$ at $z_\mathrm{min}=5.2$, motivated by observations that imply completion of reionization around $z\approx 5.2$ \citep[e.g.][]{Bosman22,Gaikwad23,Zhu24}.\footnote{We use $\epsilon=10^{-3}$ because $x_\mathrm{HI}(z)$ is modelled on the open interval $(0,1)$.} The beginning of reionization is much more uncertain. Evidence from large-scale CMB polarization and constraints on the duration of reionization from the kSZ effect \citep{Zahn12,Planck16_Reionization,Reichardt21,Smith17,SPT24_Reion} suggest that there is little contribution to $\tau$ beyond $z>15$ \citep{Planck16_Reionization,Planck20}, but see also e.g. \citet{Heinrich21,Cheng25}. Although our earliest data point is at $z=11.49$, we fix $x_\mathrm{HI}(z_\mathrm{max})=1-\epsilon$ at $z_\mathrm{max}=15$, to allow for early and gradual reionization scenarios \citep{Finkelstein19,Asthana24,Lu24,Munoz24,Cain25}. We will relax this assumption in Section~\ref{sec:extended_models}, where we consider an alternative analysis with $z_\mathrm{max}=30$ to allow for more extended reionization scenarios.

We fix the remaining two hyper parameters at $\sigma=0.3$ and $\mu(z)=\mu=0$. This choice is motivated by the behaviour at high redshifts. At low redshifts, the evolution of $x_\mathrm{HI}(z)$ is strongly constrained by the data. In the absence of constraints from the data or the boundary conditions, the results will be driven by the prior. For $z>12$, we reasonably assume that $x_\mathrm{HI}(z)\approx0$ with an uncertainty of $\sigma=0.3$.

To allow for consistent comparisons with other analyses, we require a uniform prior on $\tau$. We accomplish this with the method of \citet{Millea18} and \citet{Handley19}. Using Monte Carlo sampling, we estimate the implicit prior on $\tau$ determined by the choices outlined above. We then reweight samples accordingly in order to obtain a flat prior on $\tau$. This correction has a minimal effect on the results.

We assume that the first helium reionization happens simultaneously with hydrogen reionization and the second helium reioniziation is modelled as a smooth transition of the form \eqref{eq:tanh_form} at $z=3.5$, following \citet{Planck20}. The impact of helium reionization on the optical depth is negligible compared to the uncertainty in $\tau$ from hydrogen reionization and this remains the case for the results presented in this paper.

\subsection{Baseline results}\label{sec:baseline}

In this section, we only report results for the $\Lambda$CDM model. Fitting the reionization model to the $x_\mathrm{HI}(z)$ constraints summarized in Table~\ref{tab:data}, we obtain the reconstructed reionization histories shown in Fig.~\ref{fig:gaussian_reconstruction}. Overall, we find that the Gaussian process can describe the trends in the data well. The maximum \emph{a posteriori} realization has $\chi^2=5.4$ for $19$ data points. The reconstruction favours rapid and late reionization histories that occur mostly between $6\lesssim z\lesssim 8$. Defining the midpoint of reionization, $z_\mathrm{mid}$, as the redshift at which $x_\mathrm{HI}=0.5$ and the duration as $\Delta z_{50} = z_{75} - z_{25}$, where $z_{75}$ and $z_{25}$ are the redshifts at which $x_\mathrm{HI}=0.75$ and $x_\mathrm{HI}=0.25$, we find $z_\mathrm{mid}=6.98^{+0.13}_{-0.17}$ and $\Delta z_{50}=1.07^{+0.11}_{-0.26}$.\footnote{Technically, $z_{25},z_\mathrm{mid}$, and $z_{75}$ are not necessarily unique, but this issue only affects a minimal fraction of reconstructed histories.}

To begin to constrain the optical depth, we first approximately factor out the dependence on $H(z)$ by defining the parameter $T \equiv \tau\sqrt{\Omega_\mathrm{m}h^2 / 0.14}$.\footnote{The value $0.14$ is motivated by the result $\Omega_\mathrm{m}h^2=0.1399\pm 0.0050$ from DESI DR2 BAO and BBN.} With the addition of the BBN prior, we obtain a tight constraint,
\begin{align}
    T = 0.0470^{+0.0027}_{-0.0039} \qquad (\text{BBN + }x_\mathrm{HI}).
\end{align}

\begin{figure}
    \includegraphics[height=.3\textwidth]{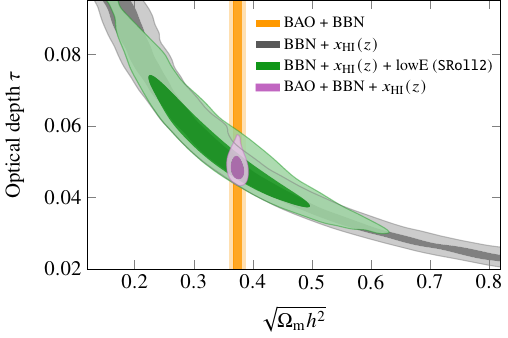}\vspace{-0.5em}
    \caption{Constraints on the optical depth, $\tau$, and the square root of the matter density, $\sqrt{\Omega_\mathrm{m}h^2}$, obtained from different combinations of DESI BAO, BBN, $x_\mathrm{HI}(z)$, and large-scale CMB polarization (lowE) data.}
    \label{fig:contours_tau}
\end{figure}

\noindent
Fig.~\ref{fig:contours_tau} shows that by adding DESI BAO data, we can break the degeneracy between $\tau$ and $\Omega_\mathrm{m}h^2$ and thereby constrain the optical depth itself. The rapid reionization histories preferred by the data then correspond to
\begin{align}
    \tau = 0.0491^{+0.0014}_{-0.0031} \qquad (\text{Baseline;\; BAO + BBN + }x_\mathrm{HI}). \label{eq:gp_tau}
\end{align}

\noindent
As an aside, we can also use $\tau$ as a ``standard depth'' and constrain $\Omega_\mathrm{m}h^2$ through the combination of $x_\mathrm{HI}(z)$ data with an independent probe of $\tau$. We demonstrate this by adding large-scale CMB polarization data using the \texttt{SRoll2} likelihood (see the next section) in Fig.~\ref{fig:contours_tau}. This combination yields
{
\setlength\arraycolsep{1.5pt}
\begin{align}
    \left.\begin{array}{ll}
        \tau & = 0.0544^{+0.0094}_{-0.014} \\
        \Omega_\mathrm{m}h^2 & = 0.132^{+0.030}_{-0.080}
    \end{array}\right\} \quad (\text{BBN + }x_\mathrm{HI}\text{ + lowE (\texttt{SRoll2})}).
\end{align}
}

\noindent
The resulting constraint on $\Omega_\mathrm{m}h^2$ is compatible with standard CMB and BAO + BBN measurements, but the error bars are not competitive. Future CMB experiments could measure $\tau$ with an uncertainty of $0.002$ \citep{LiteBIRD23}, which may increase the utility of this method.

Including BAO and BBN constraints in the likelihood helps to break parameter degeneracies, but does not significantly impact the $x_\mathrm{HI}(z)$ fit. The timing of reionization is essentially unchanged ($z_\mathrm{mid}=6.99^{+0.12}_{-0.18}$ and $\Delta z_{50}=1.09^{+0.10}_{-0.28}$). In addition to the GP reconstruction, Fig.~\ref{fig:gaussian_reconstruction} also shows some representative reionization histories that are more gradual and achieve larger optical depths, $\tau=0.07$ or $\tau=0.09$, which could alleviate or resolve the tension between DESI BAO and CMB data. These histories are inconsistent with the Lyman-$\alpha$ constraints, particularly around $7\lesssim z\lesssim 8$.

\begin{figure}
    \includegraphics[height=.3\textwidth]{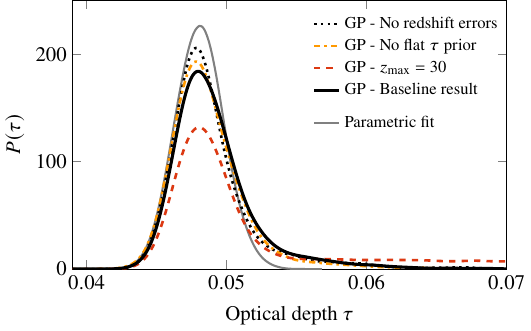}\vspace{-1.7em}
    \caption{Posterior distributions for $\tau$, obtained from the $x_\mathrm{HI}(z)$ constraints in Table~\ref{tab:data}, combined with BBN and BAO data. In addition to our baseline Gaussian process result (solid line), we show the result for the parametric fit of Eq.~\eqref{eq:tanh_form} (thin line), and variations of the Gaussian process methodology: neglecting redshift errors (dotted), not imposing a flat $\tau$ prior (dot-dashed), and increasing the maximum starting redshift to $z_\mathrm{max}=30$ (short dashed).}
    \label{fig:gaussian_tau}
\end{figure}

The resulting one-dimensional marginalized posterior distribution of $\tau$ is shown as a thick solid line in Fig.~\ref{fig:gaussian_tau}. The figure also shows the impact of various modelling assumptions. Neglecting redshift errors has a negligible impact, as does the flat-prior correction. When comparing our Gaussian process reconstruction with a parametric fit based on Eq.~\eqref{eq:tanh_form}, the results are in good agreement. For the parametric fit, we find
\begin{align}
    \tau = 0.0481\pm 0.0018 \qquad (\text{parametric;\; BAO + BBN + }x_\mathrm{HI}),
\end{align}

\noindent
compared to Eq.~\eqref{eq:gp_tau} for the GP reconstruction. Although the results are similar, the posterior distribution for the GP has a broader tail at large optical depths. This can be traced to uncertainty in the reconstruction at high $z$. Unlike the parametric form, the GP reconstruction permits an early drop and rise in the neutral hydrogen fraction before $z>9$, allowing larger values of $\tau$.

Our baseline implementation of the GP reconstruction forces the time derivative of $x_\mathrm{HI}(z)$ to vanish at the fixed endpoints of reionization. We have verified that this does not significantly impact the results by testing an alternative implementation in which the time derivative asymptotes to zero. This is done by fitting an exponential at $z < 6$, while leaving the model unchanged at higher redshifts. The effect on the optical depth is below $1\%$.

We also consider an alternative version of the highest-redshift data point in our sample. \citet{CurtisLake23} present a second fit to the same spectrum that corresponds to a higher-redshift solution at $z=11.6$, which is consistent with a significantly higher ionization fraction. Replacing the $z=11.49$ data point with the alternative at $z=11.6$, we obtain
\begin{align}
    \tau = 0.0496^{+0.0012}_{-0.0037} \quad (\text{BAO + BBN + }x_\mathrm{HI}\text{;\; alternative CL+23}),
\end{align}

\noindent
which represents only a minor shift from the baseline result. This suggests that the large error bars are sufficient to minimize the impact of this uncertainty. Indeed, the result is nearly identical when the \citet{CurtisLake23} point is excluded altogether:
\begin{align}
    \tau = 0.0497^{+0.0012}_{-0.0038} \quad (\text{BAO + BBN + }x_\mathrm{HI}\text{;\; no CL+23}).
\end{align}

\noindent
Finally, we consider an alternative compilation of $x_\mathrm{HI}(z)$ constraints from \citet{Tang24}, obtained from a comprehensive analysis of the Lyman-$\alpha$ equivalent width distribution between $6.5<z<13.3$ based on JWST data. These constraints are included as red squares in Fig.~\ref{fig:gaussian_reconstruction}. The figure shows that the constraints are consistent with the main sample, though favouring marginally lower neutral fractions around $z\approx9$. Replacing all data points in our main sample at $z>6.5$ with the \citet{Tang24} constraints, we obtain
\begin{align}
    \tau = 0.0533^{+0.0036}_{-0.0055} \quad (\text{BAO + BBN + }x_\mathrm{HI}\text{;\; Tang+24 data}). \label{eq:tang_constraint}
\end{align}

\noindent
This is slightly larger than the baseline result of Eq.~\eqref{eq:gp_tau}, though still consistent within $1\sigma$. The increase in $\tau$ is likely driven by the constraint at $z\approx9$. Compared to the baseline result, the uncertainty of Eq.~\eqref{eq:tang_constraint} is larger by a factor of two due to the smaller number of data points and the corresponding loss of granularity in redshift. See Appendix~\ref{sec:tang_fit} for further details.

\subsection{Extended reionization histories}\label{sec:extended_models}

The most important modelling assumption relates to the boundary condition at $z_\mathrm{max}=15$. Increasing the maximum starting redshift of reionization to $z_\mathrm{max}=30$ opens up the possibility of extended reionization histories. Due to the lack of observations beyond $z>13$, the possibility that there is a long tail with a non-negligible ionization fraction at high redshifts, possibly driven by Population III (Pop III) stars \citep{Qin20,Wu21}, cannot be excluded using current $x_\mathrm{HI}(z)$ constraints alone.

\begin{figure}
    \includegraphics[width=.48\textwidth]{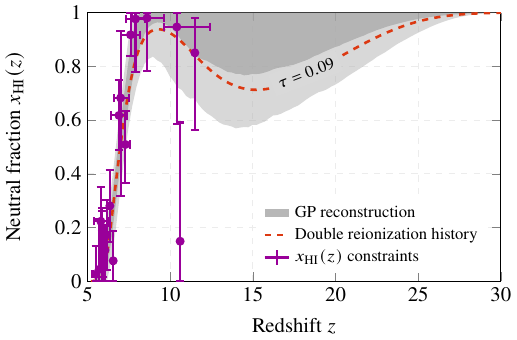}
    \caption{Gaussian process reconstruction of the reionization history using constraints from Table~\ref{tab:data}, but allowing the starting redshift of reionization to be as high as $z_\mathrm{max}=30$. The dark and light regions represent the 68\% and 95\% posterior predictive distributions. The dashed line shows a representative double reionization history, which achieves an optical depth of $\tau=0.09$ and fits the $x_\mathrm{HI}(z)$ data.
    }
    \label{fig:z30_reconstruction}
\end{figure}

The outcome of the GP reconstruction, when setting $z_\mathrm{max}=30$, is shown in Fig.~\ref{fig:z30_reconstruction}. The dashed line represents a double reionization history, which is still allowed by the $x_\mathrm{HI}(z)$ data and produces a large optical depth of $\tau=0.09$. The posterior distribution of $\tau$ in this case, shown as a short dashed line in Fig.~\ref{fig:gaussian_tau}, still peaks at the same value as the baseline case, but skews towards larger values of $\tau$. We find
\begin{align}
    \tau = 0.0480^{+0.0091}_{-0.0039} \qquad (z_\mathrm{max}=30\text{; BAO + BBN + }x_\mathrm{HI}),
\end{align}

\noindent
where we report the mode of the posterior distribution, instead of the mean, $\bar\tau=0.0583$, because it lies outside the $68\%$ credible interval.

Models with high-$z$ contributions to $\tau$ can in principle be constrained when Lyman-$\alpha$ observations are combined with kSZ or large-scale CMB polarization data \citep[e.g.][]{Qin20,Wu21,Cain25b}. Recent analyses of the kSZ effect have produced constraints on the midpoint and duration of reionization. \citet{SPT24_Reion} report an upper limit on the duration of reionization, $\Delta z_{50}<4.5$ (95\%), based on a measurement of the kSZ trispectrum  using SPT data, when imposing a prior that reionization ends by $z=6$. This is similar to the constraint $\Delta z_{50}<4.1$ (95\%) from the kSZ power spectrum analysis of \citet{Reichardt21}. However, these measurements still rely on specific assumptions about the reionization morphology and the contributions of foregrounds. A more general treatment of these uncertainties would likely weaken the constraints on $\Delta z_{50}$. For instance, the new analysis of the kSZ power spectrum from SPT \citep{Chaubal26} finds significantly different constraints on $\Delta z_{50}$, depending on which reionization simulation is adopted to make predictions. Moreover, these analyses do not yet account for all possible systematics, such as contamination by extragalactic CO emission \citep{Mehta25}.

Based on the \textsc{amber} simulations \citep{Trac22,Chen23}, the average reionization signal in the kSZ trispectrum, on scales $50\leq L \leq 300$, can be modelled as \citep{SPT24_Reion}
\begin{align}
    LC_L^{KK} = 2.73\times10^{-5}\left[\frac{\Delta z_{90}}{4.0}\right]^{1.74} \left[\frac{z_\mathrm{mid}}{8.0}\right]^{2.52}, \label{eq:kSZ_spectrum}
\end{align}

\noindent
where $\Delta z_{90} = z_{95} - z_{5}$, is defined similarly as $\Delta z_{50}$. As a first approximation, we performed the same non-parameteric reconstruction as above, with $z_\mathrm{max}=30$, but imposed a joint prior on the midpoint and duration of reionization, $p(\Delta z_{90}, z_\mathrm{mid}) = p(L C_L^{KK})$, by sampling from the joint two-dimensional marginalized posterior distribution, reproduced from Fig.~5 in \citet{SPT24_Reion}. Consistently with \citet{Cain25b}, we found that large optical depths, $\tau=0.09$, could be ruled out at more than $2\sigma$ when Lyman-$\alpha$ constraints on the beginning of reionization are combined with a kSZ prior on its duration, even when allowing $z_\mathrm{max}=30$, excluding models such as the double reionization history shown by the dashed line in Fig.~\ref{fig:z30_reconstruction}. However, we do not show the results of this analysis, because of the aforementioned uncertainties in the kSZ measurements and because Eq.~\eqref{eq:kSZ_spectrum} implicitly assumes a monotonic reionization history, which is not necessarily the case in our GP reconstruction. Moreover, it has been pointed out that Eq.~\eqref{eq:kSZ_spectrum} does not capture the reionization signal from early times, when the size of the $\mathrm{HII}$ regions is smaller than the CMB beam scale \citep{Ferraro18}.

In future, we expect that extended reionization models can be constrained by the combination of Lyman-$\alpha$ and kSZ data, since the kSZ signal is expected to be significantly larger for double reionization models \citep{McQuinn05}, but more work is needed to obtain reliable constraints.

\begin{figure*}
   \includegraphics[width=.49\textwidth]{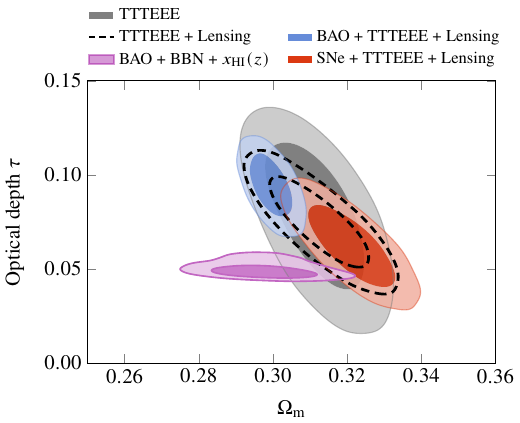}
   \includegraphics[width=.49\textwidth]{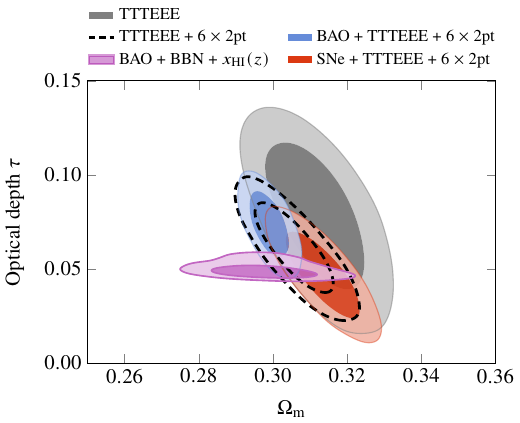}
   \caption{Constraints on the optical depth, $\tau$, and the matter fraction, $\Omega_\mathrm{m}$, in $\Lambda$CDM for various data combinations: small-scale \emph{Planck} data alone (gray) or combined with CMB lensing (black dashed). Additionally including DESI BAO (blue) or DES SNe (red) further tightens the constraints. The results from the $x_\mathrm{HI}(z)$ dataset, combined with BAO and BBN, is shown in purple. In the right panel, we also include galaxy clustering, galaxy lensing, and the cross correlation with CMB lensing, indicated as $6\times2$pt. This combination yields lower values of $S_8\equiv\sigma_8(\Omega_\mathrm{m}/0.3)^{0.5}$ and leads to better agreement among the different probes.}
   \label{fig:contours}
\end{figure*}

\subsection{Comparison with the literature}

Most previous works included a CMB optical depth constraint as input, whereas we have deliberately excluded it. However, anticipating the discussion in the next section, we find that astrophysical constraints on $x_\mathrm{HI}(z)$ lead to very tight constraints on the reionization history that agree with constraints from large-scale CMB polarization measurements. These findings are consistent with previous works \citep[e.g.][]{Greig17,Mason19b,Paoletti25,Sims25}.

Our baseline constraint, $\tau=0.0491^{+0.0014}_{-0.0031}$, is consistent with most previous analyses, but towards the lower end of the range. For instance, \citet{Mason19b} found $\tau=0.053\pm0.004$, based on a non-parametric reconstruction from Lyman-$\alpha$ data combined with a CMB optical depth constraint. We can also compare with the recent analysis of \citet{Paoletti25}. They found $\tau=0.0515_{-0.0028}^{+0.0014}$ from CMB data combined with $x_\mathrm{HI}(z)$ constraints, which is fully consistent with our result. When also including UV luminosity function data, they obtain an increased optical depth of $\tau=0.0542_{-0.0028}^{+0.0017}$, although this result is more model dependent. Another recent analysis by \citet{Sims25} found $\tau=0.052_{-0.0018}^{+0.0016}$ from a joint analysis of CMB, Lyman-$\alpha$ and Lyman-$\beta$, and 21-cm constraints. In all cases, the error bars are much smaller than the typical uncertainty from the CMB, in agreement with our findings.

The slight decrease in $\tau$ compared to previous works could be explained by the inclusion of the latest $x_\mathrm{HI}(z)$ constraints at $z>8$. In particular, the data points from \citet{Umeda25} help to pinpoint the beginning of reionization, reinforcing the preference for a late start and a low optical depth. Another contributing factor could be that the large-scale CMB polarization data that we excluded prefer slightly larger values of $\tau$, though this depends on the implementation of the likelihood (see Section~\ref{sec:lowE}).

\section{Constraints from the CMB}\label{sec:cmb}

In this section, we will consider CMB constraints on $\tau$. Since these data are less sensitive to the shape of the reionization history, we adopt the simpler parametric form of Eq.~\eqref{eq:tanh_form} for all CMB-based analyses that exclude direct $x_\mathrm{HI}(z)$ constraints. This allows us to explore a greater variety of models at reduced computational cost.

\subsection{\boldmath Results for $\Lambda$CDM}

Let us begin by discussing the constraints from small-scale CMB data and then move on to large-scale CMB polarization.

\subsubsection{Small-scale measurements}\label{sec:small_scale_cmb}

It is interesting to consider what constraints on $\tau$ are possible using only small-scale CMB measurements \citep{Giare24b,Sailer25,Jhaveri25}. When analyzing such data within $\Lambda$CDM, there is a strong degeneracy between the parameters $\tau$ and $\Omega_\mathrm{m}$. This is especially true when including information from CMB lensing \citep{Jhaveri25}, as may be gleaned from the left panel of Fig.~\ref{fig:contours}. The solid gray contours represent the constraints from small-scale \emph{Planck} data alone. The black dashed contours show that these constraints are significantly tightened, in the direction orthogonal to the $\Omega_\mathrm{m}$-$\tau$ degeneracy, by adding the CMB lensing likelihood of \citet{ACT_SPT_Planck_Lensing}, which is based on \emph{Planck}, ACT, and SPT data. This can be explained by the fact that CMB lensing measurements are sensitive to $\sigma_8\Omega_\mathrm{m}^{0.25}$, the previously mentioned degeneracy between $\tau$ and $A_\mathrm{s}$, and the dependence of the clustering amplitude, $\sigma_8$, on the amplitude of the primordial power spectrum, $A_\mathrm{s}$.

A similar conclusion holds when we add low-redshift galaxy clustering and lensing measurements, which are sensitive to $\sigma_8\Omega_\mathrm{m}^{0.5}$. We demonstrate this in the right panel of Fig.~\ref{fig:contours}, where we add the $6\times2$pt likelihood of \citet{Abbott23}, which is based on DES, \emph{Planck}, and SPT measurements. The degeneracy between $\Omega_\mathrm{m}$ and $\tau$ is again apparent from the black dashed contours, although with an expected rotation in the direction of the degeneracy.

To make progress, the matter density fraction can be further constrained using low-redshift probes. DESI BAO precisely determine $\Omega_\mathrm{m} = 0.2975\pm 0.0086$ \citep{DESI25.II}. The blue contours in the left panel of Fig.~\ref{fig:contours} show that the combination of small-scale \emph{Planck} data, CMB lensing, and DESI BAO yields a large value of $\tau=0.094\pm0.011$. These constraints are in tension with those from BAO, BBN, and $x_\mathrm{HI}(z)$ data, which were discussed extensively in Section~\ref{sec:nonpar} and are shown in purple in Fig.~\ref{fig:contours}. The DES compilation of type 1a SNe prefers a significantly larger $\Omega_\mathrm{m}=0.352\pm0.017$ \citep{DES24}. Moving along the CMB degeneracy, this higher $\Omega_\mathrm{m}$ gives rise to a smaller $\tau=0.062\pm0.014$, as shown by the red contours. This constraint is compatible with the $x_\mathrm{HI}(z)$ measurement of $\tau$. However, the tension in $\Omega_\mathrm{m}$ between BAO and SNe is also apparent between the purple and red contours.

In the right panel of Fig.~\ref{fig:contours}, we see better agreement among the different probes when including the $6\times2$pt likelihood of \citet{Abbott23}. This can be explained by the fact that the $6\times2$pt analysis yields lower values of $S_8\equiv\sigma_8(\Omega_\mathrm{m}/0.3)^{0.5}=0.792\pm0.012$ \citep{Abbott23}, compared to $S_8 = 0.828 \pm 0.012$ for the CMB lensing analysis of \citet{ACT_SPT_Planck_Lensing}. As a result, the combinations with DESI BAO and DES SNe yield smaller values of $\tau$ that are more compatible with the $x_\mathrm{HI}(z)$ constraint. In the case of DES SNe, we find $\tau=0.047\pm0.015$, with a central value close to that of the $x_\mathrm{HI}(z)$ analysis. It is important to note that, despite the widely reported tension in $S_8$ \citep[e.g.][]{Abdalla22,DiValentino25}, the results of the $6\times$2pt analysis of \citet{Abbott23} are consistent with the primary \emph{Planck} CMB measurements (see also \citealt{Chen24,DESKiDS23,DES24b,DESI24.KP7B,Wright25}), justifying their combination.

In summary, while it is possible to obtain tight constraints on $\tau$ by combining small-scale CMB data with low-redshift probes, the degeneracy between $\tau$ and other cosmological parameters, particularly $\Omega_\mathrm{m}$ and $\sigma_8$, means that the results are sensitive to the choice of low-redshift probe. Later on, we will see that the results are also sensitive to the choice of cosmological model.

\begin{table}
    {
    \renewcommand{\arraystretch}{1.425}
    \begin{center}
        \caption{Constraints on the reionization optical depth, $\tau$, from small-scale CMB data combined with various low-redshift probes as indicated, from large-scale CMB polarization data (lowE) for four different likelihoods, and from $x_\mathrm{HI}(z)$ data. In all cases, we assume the $\Lambda$CDM model.}
        \label{tab:xHI_lowE}
        \begin{tabular}{lc}
            \hline
            Dataset & $\tau$ \\
            \hline
            SNe + TTTEEE + $6\times2$pt & $0.047\pm 0.015$\\
            SNe + TTTEEE + Lensing & $0.062\pm 0.014$\\
            BAO + TTTEEE + $6\times2$pt & $0.073\pm 0.012$ \\
            BAO + TTTEEE + Lensing & $0.094\pm 0.011$ \\
            BAO + ACT-SPT + Lensing & $0.094\pm 0.011$ \\
            BAO + P-ACT-SPT + Lensing & $0.0914\pm 0.0094$ \\
            \hline
            BAO + BBN + lowE (\texttt{SimAll}) & $0.058^{+0.010}_{-0.018}$ \\
            BAO + BBN + lowE (\texttt{SRoll2}) & $0.0535^{+0.0066}_{-0.011}$ \\
            BAO + BBN + lowE (\texttt{LoLLiPoP}) & $0.0494^{+0.0059}_{-0.0082}$ \\
            BAO + BBN + lowE (\texttt{ELiCA}) & $0.0508^{+0.0057}_{-0.0075}$ \\
            \hline
            BAO + BBN + $x_\mathrm{HI}(z)$ & $0.0491^{+0.0014}_{-0.0031}$ \\
            \hline
        \end{tabular}
    \end{center}
    }
\end{table}

\subsubsection{Large-scale measurements}\label{sec:lowE}

\begin{figure}
    \includegraphics[width=.5\textwidth]{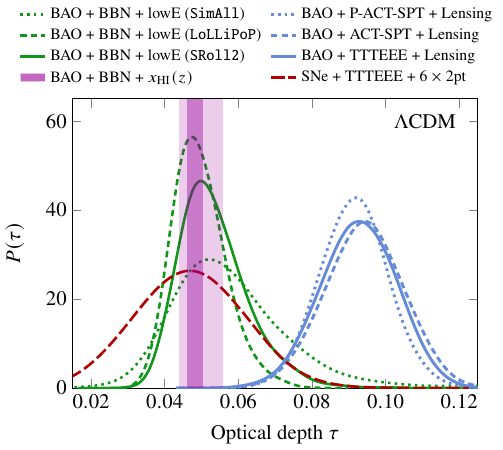}
    \caption{One-dimensional marginalized constraints on the optical depth, $\tau$, from a wide range of data combinations for $\Lambda$CDM. The constraints from our non-parametric reconstruction of the reionization history are shown as vertical shaded bands ($1\sigma$ and $2\sigma$ bounds). Also shown are the constraints from large-scale CMB polarization for three different likelihoods (dark green), the combination of small-scale CMB anisotropies with BAO for three different CMB datasets (light blue), and the combination of small-scale CMB with type 1a supernovae and $6\times2$pt clustering measurements (red long dashed).}
    \label{fig:tau_results}
\end{figure}

Let us move on to the constraints from large-scale CMB polarization (lowE). We consider four different likelihoods based on \emph{Planck} data: \texttt{SimAll} \citep{Planck20_Spectra}, \texttt{SRoll2} \citep{Delouis19}, \texttt{LoLLiPoP} \citep{Tristram21,Tristram24}, and \texttt{ELiCA} \citep{Genesini26}. Compared to the \texttt{SimAll} likelihood used in the Planck 2018 analysis \citep{Planck20}, the \texttt{SRoll2} likelihood uses an improved version of the \texttt{SRoll} mapmaking algorithm that corrects for known instrumental effects. Both \texttt{SimAll} and \texttt{SRoll2} use an empirical probability distribution constructed from simulated power spectra. \texttt{LoLLiPoP} uses a Gaussian likelihood function with a covariance matrix based on simulated spectra, handling non-Gaussianities with an approximation based on \citet{Hamimeche08}. Finally, we consider the new \texttt{ELiCA} likelihood from \citet{Genesini26}, which also relies on the \texttt{SRoll2} mapmaking algorithm and the \citet{Hamimeche08} approximation, while including additional data from WMAP \citep{Natale20}.

Although it is possible to constrain $\tau$ using lowE data alone, even when varying all $\Lambda$CDM parameters, we combine these likelihoods with DESI BAO and BBN constraints to break parameter degeneracies, which improves the constraint on $\tau$ by $40\%$. For instance, in the case of \texttt{SRoll2}, we find
\begin{align}
    \tau = 0.0526^{+0.0081}_{-0.015}\qquad (\text{lowE (}\texttt{SRoll2}\text{)}),
\end{align}

\noindent
while adding BBN and BAO improves this to 
\begin{align}
    \tau = 0.0535^{+0.0066}_{-0.011}\qquad (\text{BAO + BBN + lowE (}\texttt{SRoll2}\text{)}).
\end{align}

\noindent
The results for the other three likelihoods are given in Table~\ref{tab:xHI_lowE}. The four likelihoods are mutually consistent and in good agreement with the Lyman-$\alpha$ constraint of Section~\ref{sec:nonpar}. Compared to \texttt{SimAll}, the newer likelihoods produce tighter constraints that shift closer to the Lyman-$\alpha$ bound.

\begin{table*}
    {
    \renewcommand{\arraystretch}{1.425}
    \begin{center}
        \caption{Constraints for different cosmological models on the reionization optical depth, $\tau$, and the parameter combination $T=\tau\sqrt{\Omega_\mathrm{m}h^2/0.14}$ that is more directly measured by $x_\mathrm{HI}(z)$ constraints. The results are given as 68\% credible intervals for four different data combinations labelled 1A through 3. The final column lists the tension in $T$ between combinations 1A ($x_\mathrm{HI}(z)$ data with a BBN prior) and 2 (constraints from BAO, \emph{Planck} TTTEEE, and CMB lensing).}
        \label{tab:tensions}
        \begin{tabular}{lccccccc}
            \hline
            & 1A. & 1B. & \multicolumn{2}{c}{2.} &  \multicolumn{2}{c}{3.} & Tension in $T$  \\
            & BBN + $x_\mathrm{HI}(z)$ & BAO + BBN + $x_\mathrm{HI}(z)$ &  \multicolumn{2}{c}{BAO + TTTEEE + Lensing} & \multicolumn{2}{c}{BAO + BBN + lowE (\texttt{SRoll2})} & between \\
            Cosmological model & $T$ & $\tau$ & $T$ & $\tau$ & $T$ & $\tau$ & $1\mathrm{A}.\leftrightarrow2$ \\ 
            \hline
            $\Lambda$CDM & $0.0470^{+0.0027}_{-0.0039}$ & $0.0491^{+0.0014}_{-0.0031}$ &  $0.094\pm 0.011$ & $0.094\pm 0.011$ & $0.0535^{+0.0066}_{-0.011}$ & $0.0535^{+0.0066}_{-0.011}$ & $3.6\sigma$ \\
            $\Lambda$CDM $+\;\alpha_\mathrm{s}$ & $0.0470^{+0.0027}_{-0.0039}$ & $0.0491^{+0.0014}_{-0.0031}$ &  $0.093\pm 0.011$ & $0.093\pm 0.011$ & $0.0520^{+0.0078}_{-0.013}$ & $0.0520^{+0.0078}_{-0.013}$ & $4.0\sigma$ \\
            $\Lambda$CDM $+\;A_\mathrm{lens}$ & $0.0470^{+0.0027}_{-0.0039}$ & $0.0491^{+0.0014}_{-0.0031}$ &  $0.094^{+0.035}_{-0.055}$ & $0.094^{+0.036}_{-0.055}$ & $0.0546^{+0.0069}_{-0.011}$ & $0.0546^{+0.0069}_{-0.011}$ & $0.9\sigma$ \\
            $\Lambda$CDM $+\;\Omega_\mathrm{k}$ & $0.0481^{+0.0073}_{-0.0060}$ & $0.0498^{+0.0016}_{-0.0036}$ &  $0.086\pm 0.012$ & $0.085\pm 0.012$ & $0.0537^{+0.0079}_{-0.012}$ & $0.0535^{+0.0074}_{-0.011}$ & $2.7\sigma$ \\
            $\Lambda$CDM $+\;N_\mathrm{eff}$ & $0.0476\pm 0.0071$ & $0.0495^{+0.0017}_{-0.0034}$ &  $0.091\pm 0.011$ & $0.091\pm 0.012$ & $0.0537^{+0.0070}_{-0.011}$ & $0.0538^{+0.0071}_{-0.011}$ & $3.2\sigma$ \\
            $\Lambda$CDM $+\,\sum m_{\nu,\mathrm{eff}}$ & $0.0471^{+0.0028}_{-0.0041}$ & $0.0492^{+0.0030}_{-0.0047}$ &  $0.063^{+0.024}_{-0.021}$ & $0.063^{+0.024}_{-0.021}$ & $0.0542^{+0.0075}_{-0.011}$ & $0.0542^{+0.0071}_{-0.011}$ & $0.8\sigma$ \\
            $w$CDM & $0.0480^{+0.0040}_{-0.0029}$ & $0.0504^{+0.0019}_{-0.0035}$ &  $0.091^{+0.014}_{-0.012}$ & $0.091\pm 0.013$ & $0.0528^{+0.0070}_{-0.011}$ & $0.0545^{+0.0070}_{-0.012}$ & $3.0\sigma$ \\
            $w_0w_a$CDM & $0.0480^{+0.0040}_{-0.0032}$ & $0.0483^{+0.0015}_{-0.0035}$ &  $0.069^{+0.014}_{-0.016}$ & $0.069^{+0.014}_{-0.016}$ & $0.0540^{+0.0072}_{-0.011}$ & $0.0524^{+0.0069}_{-0.011}$ & $1.3\sigma$ \\
            \hline
        \end{tabular}
    \end{center}
    }
\end{table*}

In Fig.~\ref{fig:tau_results}, we present the marginalized posterior distributions of $\tau$ for the various CMB datasets we have considered, leaving out \texttt{ELiCA} for visual clarity as it is similar to \texttt{LoLLiPoP}. The other three large-scale polarization likelihoods are shown in dark green, demonstrating the consistency between \texttt{SimAll} (dotted green), \texttt{SRoll2} (solid green), and \texttt{LoLLiPoP} (dashed green). These measurements are also consistent with the $x_\mathrm{HI}(z)$ constraint, which is shown as a purple shaded band. The combination of DES SNe, small-scale \emph{Planck} data, and the $6\times2$pt likelihood of \citet{Abbott23}, represented by the red long dashed line, also agrees well with the large-scale polarization and $x_\mathrm{HI}(z)$ bounds. 

A second set of posteriors, centred on larger values of $\tau$, are shown in light blue. These constraints arise from the combination of DESI BAO, CMB lensing, and small-scale CMB data. Compared to the combination with primary CMB measurements from \emph{Planck} (solid blue), we find very similar results with the alternative ACT-SPT dataset (dashed blue) and the most powerful P-ACT-SPT combination (dotted blue). See Table~\ref{tab:xHI_lowE} for the numerical results. In all three cases, the constraints are in tension with the $x_\mathrm{HI}(z)$ bound (see below).

\subsection{Results for extended cosmological models}\label{sec:extended_cosmologies}

In this section, we consider extensions of the $\Lambda$CDM model, including variations in the shape of the primordial power spectrum, the amplitude of the lensing potential, spatial curvature, neutrino content, and dark energy parameters. Table~\ref{tab:tensions} provides constraints on $\tau$ for all cosmological models, obtained using three different approaches. The first is a non-parametric reconstruction of the reionization history, based primarily on $x_\mathrm{HI}(z)$ constraints. The second approach relies on constraints from small-scale \emph{Planck} data, combined with CMB lensing and DESI BAO data. The third approach relies primarily on large-scale \emph{Planck} polarization measurements.

We also provide constraints on $T=\tau\sqrt{\Omega_\mathrm{m}h^2/0.14}$, which can be precisely measured with $x_\mathrm{HI}(z)$ data, independently of probes of the expansion history. The final column in Table~\ref{tab:tensions} lists the tension between the constraints on $T$ from BBN and $x_\mathrm{HI}(z)$ data and those from small-scale \emph{Planck}, CMB lensing, and DESI BAO. The reason for computing the tension in $T$ is that this parameter can be constrained with $x_\mathrm{HI}(z)$ data, independently of DESI BAO. The resulting constraints are therefore uncorrelated with those from CMB and BAO data, which simplifies the calculations. We use the \texttt{tensiometer} package\footnote{\url{https://github.com/mraveri/tensiometer}} to compute the level of tension, employing the parameter shift metric of \citet{Raveri20} and \citet{Raveri21}.

Before discussing each model in detail, we first make some general remarks. We observe that there is a $3.6\sigma$ tension in $T$ when assuming $\Lambda$CDM. The tension is reduced to $\sim3\sigma$ for three models, for which the CMB and BAO constraints shift closer towards the Lyman-$\alpha$ bound. For three other models, the tension is reduced to below $2\sigma$. This is the case when the gravitational lensing potential is scaled by the $A_\mathrm{lens}$ parameter, when allowing for an effective cosmological neutrino mass parameter, $\sum m_{\nu,\mathrm{eff}}$, and when allowing for dynamical dark energy with an evolving equation of state (`$w_0w_a$CDM'). In these cases, the optical depth is consistent with but still slightly larger than the result from BBN and $x_\mathrm{HI}(z)$ data.

The most notable difference between the three approaches to constraining $\tau$ relates to their sensitivity to the choice of cosmological model. The distinctive signature of reionization on the large-scale polarization power spectrum means that the lowE constraints are largely insensitive to the choice of cosmological model. Similarly, the $x_\mathrm{HI}(z)$ constraints fix the reionization history, essentially independently of the cosmological model. By contrast, the constraints on $\tau$ from small-scale \emph{Planck}, CMB lensing, and DESI BAO data do depend on the choice of cosmology, with shifts of up to $\sim3\sigma$ from the $\Lambda$CDM value.

Some remarks are also in order about the validity of the $x_\mathrm{HI}(z)$ constraints in extended cosmological models. The dark pixel constraints are truly model-independent, but damping wing and equivalent width constraints typically rely on reionization simulations that assume a given cosmological model \citep[e.g.][]{Mesinger16}. In principle, there could be additional dependence on the cosmological model, not captured in our analysis. As a concrete example, consider variations in the luminosity distance, $D_L(z)$, to galaxies at $z\sim7-10$. Generalizing from $\Lambda$CDM to the extended cosmological models considered below, the posterior predictive mean of $D_L(z)$ shifts by $\sim1\%$ for models like $w_0w_a$CDM and by up to $\sim3\%$ for $\Lambda$CDM with curvature $\Omega_\mathrm{k}\neq 0$, given DESI BAO data. This could bias the inferred UV absolute magnitudes, $M_\mathrm{UV}$, of galaxies at these redshifts by up to $0.07\,\si{mag}$ and propagate further to the inferred neutral fractions. This effect should be quantified precisely, but is likely subdominant compared to other modelling uncertainties, such as the $0.3\,\si{dex}$ scatter in the relationship between $M_\mathrm{UV}$ and halo mass, $M_\mathrm{h}$, used in equivalent width analyses \citep{Mason18}. The coupling between cosmological parameters and galaxy formation processes could also produce subtle effects \citep[e.g.][]{Elbers25_FLAMINGO, Pranjal25}, which are similarly at the percent-level and expected to be subdominant compared to the uncertainty in subgrid reionization parameters. A more detailed investigation of such effects will be left for future work.
 
In the following subsections, we will briefly comment on each cosmological model and assess the tension in $T$. In cases where the tension between CMB, BAO, and $x_\mathrm{HI}(z)$ data is resolved, we will combine the most powerful CMB dataset (P-ACT-SPT with lensing) and the most powerful BAO dataset (DESI DR2) with a dataset that tightly constrains the optical depth (Table~\ref{tab:data}) to obtain the strongest possible constraints on extensions of $\Lambda$CDM.

\subsubsection{Running of the spectral index}

In $\Lambda$CDM, the spectral index, $n_\mathrm{s}$ is independent of scale, but inflationary models predict departures from pure scale invariance \citep{Escudero16}. The running of the spectral index is defined as $\alpha_\mathrm{s}=\mathrm{d}n_\mathrm{s}/\mathrm{d}\!\log k$. Table~\ref{tab:tensions} shows that allowing for non-zero running does not significantly impact the constraints on $\tau$. The tension between the Lyman-$\alpha$ constraint and the CMB + BAO combination remains at the $4\sigma$-level.

\subsubsection{Lensing excess}

The CMB lensing anomaly refers to the preference of CMB data for a greater amount of gravitational lensing, either measured through the lensing smoothing effect on the primary CMB measurements or through the lensing reconstruction from the four-point correlation function, than the level that would be self-consistently predicted within a given cosmological model. It is commonly quantified using the $A_\mathrm{lens}$ parameter that scales the lensing potential \citep{Calabrese08,Renzi18,Mokeddem23}.

By constraining the matter density fraction, BAO data play an important role in the $A_\mathrm{lens}$ anomaly. As discussed in Section~\ref{sec:small_scale_cmb}, CMB lensing measurements impose a strong degeneracy between $\tau$ and $\Omega_\mathrm{m}$, such that the lower $\Omega_\mathrm{m}$ measured by DESI leads to a higher $\tau$ than the one preferred by Lyman-$\alpha$ constraints or large-scale CMB polarization measurements. Introducing the $A_\mathrm{lens}$ factor breaks the link between $\Omega_\mathrm{m}$ and $\tau$, making it possible to achieve a lower $\tau$ for a given $\Omega_\mathrm{m}$ through a value $A_\mathrm{lens}$ greater than unity.

Consequently, the tension in the optical depth can be entirely absorbed by adopting $A_\mathrm{lens}>1$. The tension in $T$ is reduced to $0.9\sigma$. However, combining the CMB dataset from P-ACT-SPT with DESI BAO and the Lyman-$\alpha$ constraints then leads to
\begin{align}
    A_\mathrm{lens} = 1.085\pm 0.022\qquad (\text{CMB + BAO + }x_\mathrm{HI}),
\end{align}

\noindent
corresponding to a $3.5\sigma$ preference for $A_\mathrm{lens}>1$.

\subsubsection{Curvature}

The lensing anomaly is related to the preference of CMB data (including large-scale polarization) for a closed Universe with $\Omega_\mathrm{k}<0$, due to the correlation between $A_\mathrm{lens}$ and $\Omega_\mathrm{k}$ \citep{DiValentino20}. This can be seen as another expression of the preference of small-scale CMB data for an optical depth greater than the one allowed by lowE or Lyman-$\alpha$ constraints.

Allowing for non-zero spatial curvature has some impact on the CMB + BAO constraint on $\tau$, shifting the best-fitting value by $0.8\sigma$ towards lower optical depths (see Table~\ref{tab:tensions}). However, $\Omega_\mathrm{k}<0$ is no longer preferred once BAO data are included \citep{Chen25}. Indeed, the combination of DESI and CMB data is consistent with a flat Universe \citep{DESI24.KP7A}. Allowing for $\Omega_\mathrm{k}\neq0$ therefore only reduces the tension with the Lyman-$\alpha$ bound to $2.7\sigma$. 

\subsubsection{New relativistic species}

We also consider the possibility of additional light degrees of freedom, as parametrized by the effective number of neutrino species, $N_\mathrm{eff}$. The Standard Model prediction for the non-instantaneous decoupling of three neutrinos is $N_\mathrm{eff}=3.044$ \citep{Froustey20,Bennett21,Binder24}. The data are fully consistent with the Standard Model prediction assumed in our baseline analysis. Consequently, allowing variations in $N_\mathrm{eff}$ does not significantly impact the constraints on $\tau$ and the tension in $T$ remains at $3.2\sigma$.

\begin{figure}
    \includegraphics[width=.5\textwidth]{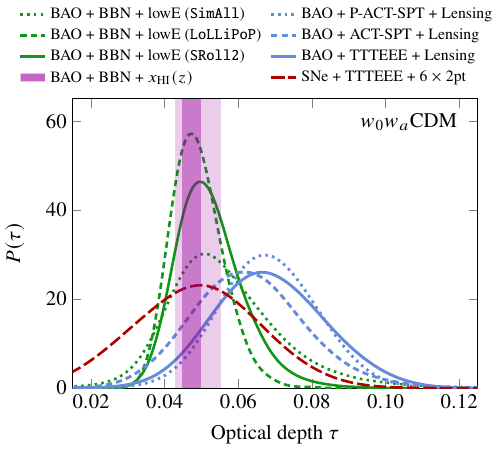}
    \caption{One-dimensional marginalized constraints on the optical depth, $\tau$, from a wide range of data combinations for $w_0w_a$CDM. The constraints from our non-parametric reconstruction of the reionization history are shown as vertical shaded bands ($1\sigma$ and $2\sigma$ bounds). Also shown are the constraints from large-scale CMB polarization for three different likelihoods (dark green), the combination of small-scale CMB anisotropies with BAO for three different CMB datasets (light blue), and the combination of small-scale CMB with type 1a supernovae and $6\times2$pt clustering measurements (red long dashed).}
    \label{fig:tau_results_w0wa}
\end{figure}

\subsubsection{Neutrino masses}

One of the most striking anomalies arising from the tension between CMB and BAO data is the cosmological preference for negative effective neutrino masses \citep{Craig24,Green25,Elbers25,Elbers_DESI25}. As pointed out in these and other works \citep[e.g.][]{Loverde24,Lynch25}, this preference is also related to the $A_\mathrm{lens}$ and $\tau$ problems.

To investigate this anomaly, we adopt the effective neutrino mass parametrization of \citet{Elbers25}, which allows for negative neutrino contributions to the energy density at late times, implemented by extending standard perturbative calculations. This mathematical extension allows us to test for signs of new physics or systematics that correlate with the cosmological neutrino mass parameter, $\sum m_{\nu,\mathrm{eff}}$. When we allow $\sum m_{\nu,\mathrm{eff}}<0$, the tension in $T$ is greatly reduced from $3.6\sigma$ to $0.8\sigma$. A significant part of this reduction is due to a $3\sigma$ shift of the best-fitting value of $\tau$ for the CMB + BAO combination towards lower optical depths. This justifies the combination of CMB, BAO, and Lyman-$\alpha$ constraints on reionization. For the most constraining CMB dataset that combines small-scale P-ACT-SPT data with CMB lensing, we find
\begin{align}
    \textstyle\sum m_{\nu,\mathrm{eff}} = -0.120\pm 0.038\,\si{\eV}\qquad (\text{CMB + BAO + }x_\mathrm{HI}). \label{eq:mnu_eff}
\end{align}

\noindent
The marginalized error is $\sigma(\sum m_{\nu,\mathrm{eff}})=0.038\,\si{\eV}$. To our knowledge, this is the strongest cosmological bound on the neutrino mass, representing a $30\%$ improvement in precision compared to \citet{Elbers_DESI25}. However, much like the extension with $A_\mathrm{lens}$, we have absorbed the tension between the underlying datasets into a new parameter that takes unphysical values. The tension with the minimal neutrino mass under the normal ordering, $\sum m_\nu>\SI{0.059}{\eV}$ \citep{Esteban24}, reaches $3.6\sigma$. Physically, the finding that $\sum m_{\nu,\mathrm{eff}}<0$ indicates a preference in the data for enhanced small-scale clustering and a late-time reduction in the matter density fraction, $\Omega_\mathrm{m}$, which represent a reversal of the cosmological effects of standard massive neutrinos. The connection with $\tau$ is then apparent from Fig.~\ref{fig:contours}.

\subsubsection{Dark energy}

As a final extension, we consider variations of the dark energy model. Generalizing from $\Lambda$CDM, we first consider dynamical dark energy with a constant equation of state, $w=P/\rho$, where $w$ is a free parameter. Adopting this `$w$CDM' model does not significantly shift the constraints on $\tau$. The tension between the Lyman-$\alpha$ bound and the BAO and CMB constraints remains at $3\sigma$. This is consistent with expectations, given that current BAO and CMB data do not prefer $w$CDM over $\Lambda$CDM \citep[e.g.][]{DESI25.II}.

Next, we consider models of dynamical dark energy with an evolving equation of state, described by the Chevallier-Polarski-Linder parametrization \citep{Chevallier01,Linder03}, in which
\begin{align}
    w(a) = w_0 + w_a(1-a),
\end{align}

\noindent
where $w_0$ and $w_a$ are free parameters. BAO measurements from the first data release of DESI \citep{DESI24.KP7A} were consistent with $\Lambda$CDM, but preferred $w_0w_a$CDM when combined with CMB and SNe data. With the second data release of DESI, this preference increased to $2.8\sigma$-$4.2\sigma$, depending on the SNe dataset \citep{DESI25.II}. See \citet{Lodha25} for an extended dark energy analysis of the DESI results and \citet{Giare24} for references to the recent literature.

Allowing for an evolving equation of state of dark energy allows the BAO and CMB combination to be more compatible with lower optical depths. The tension in $T$ is reduced to $1.3\sigma$. Fig.~\ref{fig:tau_results_w0wa} shows the marginalized posterior distributions of $\tau$ for various data combinations, assuming $w_0w_a$CDM. It is fruitful to compare this figure with its $\Lambda$CDM counterpart (Fig.~\ref{fig:tau_results}). First of all, we observe that the posteriors for the large-scale polarization measurements (dark green lines) are very similar to ones obtained under $\Lambda$CDM, demonstrating their robustness to changes in the assumed cosmological model. Both the Lyman-$\alpha$ bound (purple band) and the posterior from the combination of CMB, SNe, and $6\times2$pt measurements (red long dashed line) are also similar to their $\Lambda$CDM counterparts, but with larger uncertainties. By contrast, the posteriors from the combinations of DESI BAO, CMB, and CMB lensing data (light blue lines) are shifted towards lower $\tau$, such that all constraints are compatible with a rapid and late reionization history.

\begin{figure}
    \includegraphics[width=.48\textwidth]{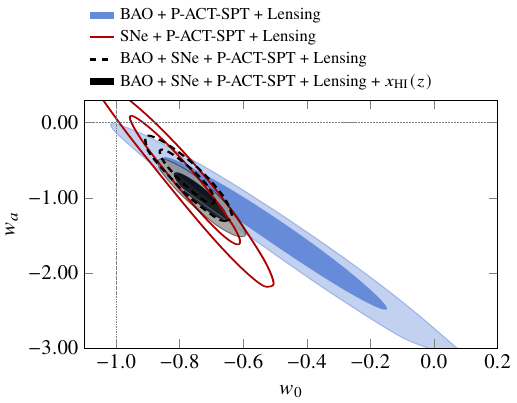}\vspace{-1em}
    \caption{Constraints on the dark energy equation of state parameters for various data combinations. Shown are the constraints from CMB and CMB lensing, combined with BAO (light blue) or SNe (open red), as well as the combination of all four (black dashed) and the combination of all four with the addition of $x_\mathrm{HI}(z)$ data (filled black).}
    \label{fig:extension_plots}
\end{figure}

Given that the tension in $T$ is significantly reduced, we combine CMB, BAO, and $x_\mathrm{HI}(z)$ data to constrain the dark energy parameters. We also include the DES SNe dataset, since the constraints from SNe and BAO are compatible within $w_0w_a$CDM (see Fig.~\ref{fig:extension_plots}). Using the CMB dataset from P-ACT-SPT, we obtain the marginalized constraints
{
\setlength\arraycolsep{1.5pt}
\begin{align}
    \left.\begin{array}{ll}
        w_0 &= -0.731\pm 0.056\\
        w_a &= -0.99^{+0.22}_{-0.20}
    \end{array}\right\}\quad (\text{CMB + BAO + SNe + }x_\mathrm{HI}), \label{eq:w0wa_results}
\end{align}
}

\noindent
which are close to the values reported in \citet{DESI24.KP7A} for the combination of DESI BAO, CMB, and DES SNe. Fig.~\ref{fig:extension_plots} illustrates how the addition of $x_\mathrm{HI}(z)$ data pulls the constraints away from $\Lambda$CDM. We estimate the preference for $w_0w_a$CDM to be $4.3\sigma$ when combining P-ACT-SPT, DESI BAO, DES SNe, and $x_\mathrm{HI}(z)$ data.\footnote{We employ the $\Delta \chi^2_\mathrm{MAP}$ metric used by \citet{DESI24.KP7A,DESI25.II}.} This is a small increase compared to the $4.2\sigma$ preference reported in \citet{DESI25.II}. Compared to that analysis, we have replaced their CMB dataset with the more constraining P-ACT-SPT dataset (including lensing), but excluded large-scale CMB temperature and polarization data, which we replaced with our $x_\mathrm{HI}(z)$ likelihood. Including large-scale temperature and polarization data would likely further increase the preference for $w_0w_a$CDM.

Since we deliberately excluded lowE data, these results are invulnerable to any large-scale CMB systematics. On the other hand, one must now contend with the possibility of systematics in the analysis of Lyman-$\alpha$ damping wings and equivalent widths, particularly at the highest redshifts \citep{Mesinger08,Keating24,Umeda24,Huberty25}.

\section{Conclusion}\label{sec:discussion}

Motivated by anomalies in the latest cosmological data, we have presented new constraints on the reionization optical depth, $\tau$, using several independent methods. Our strongest bound derives from a non-parametric reconstruction of the reionization history from Lyman-$\alpha$ and Lyman-$\beta$ constraints on the evolution of the volume-averaged neutral hydrogen fraction, $x_\mathrm{HI}(z)$, between $5<z<12$. The constraints mainly derive from dark pixel fraction constraints at low redshifts and damping wing and equivalent width constraints at high redshifts (see Table~\ref{tab:data} for references).

Our results point to a late and rapid reionization history, with a midpoint at $z_\mathrm{mid}=6.99^{+0.12}_{-0.18}$ and a duration of $\Delta z_{50}=1.09^{+0.10}_{-0.28}$. We obtain an optical depth, $\tau$, that is perfectly consistent with \emph{Planck} measurements of the large-scale CMB polarization power spectrum. The strong constraining power of quasar damping wings and equivalent widths and the consistency between damping wing, equivalent width, dark pixel fraction, and CMB constraints on the optical depth agree with the findings of earlier works \citep[e.g.][]{Greig17}. Our baseline constraint, $\tau=0.0491^{+0.0014}_{-0.0031}$, is consistent with previous analyses \citep{Mason19b,Paoletti25,Sims25}, but incorporates the latest data and is independent of the CMB.

In our baseline analysis, we assume that the onset of reionization occurs no earlier than $z=15$. When the starting redshift is increased to $z=30$, we obtain a posterior mode of $\tau = 0.0480^{+0.0091}_{-0.0039}$, close to the baseline result, but with a long tail towards higher values of $\tau$, thereby shifting the mean to $\bar\tau=0.0583$. In this case, the posterior depends more strongly on the prior, due to the lack of observations at $z>12$.

Under the assumption of the cosmological $\Lambda$CDM model, BAO data from DESI, combined with CMB lensing and small-scale CMB data from \emph{Planck} (or the independent ACT and SPT datasets) imply a larger optical depth, $\tau=0.094\pm0.011$. Such a large optical depth provides a solution to the tension between DESI BAO and CMB constraints that preserves the $\Lambda$CDM model \citep{Sailer25,Jhaveri25}. However, it would have a number of consequences that we summarize below:
\begin{itemize}
    \item The large optical depth is in $3.6\sigma$ tension with our Lyman-$\alpha$ bound, unless we relax the assumption about the maximum starting redshift of reionization and allow for reionization scenarios with a significant contribution to $\tau$ from $z>15$, as in models in which the IGM is reionized multiple times \citep{Cen03,Furlanetto05,Mason19b,Ahn21,Tan25} or with a long high-redshift tail from Pop III stars \citep{Qin20,Wu21}.
    \item We found that such early reionization models are still disfavoured when including a prior on the midpoint and duration of reionization based on recent measurements of the kSZ effect \citep{SPT24_Reion,Cain25b}, but caution that these measurements are still subject to large uncertainties and implicitly assume a monotonic reionization history. Further work is needed to study the kSZ effect in non-monotonic models.
    \item It would also conflict with \emph{Planck} lowE measurements. There is no evidence for systematic errors in these measurements, but \citet{Giare24b} give some reasons why the possibility cannot be dismissed. \citet{Jhaveri25} offer another way of relaxing the lowE constraint on $\tau$, through a change in the curvature fluctuation power spectrum on large scales. However, this would no longer be a standard $\Lambda$CDM scenario.
    \item A large optical depth would also be in tension with the constraint from the combination of small-scale CMB data with type 1a supernovae and $6\times2$pt galaxy lensing, galaxy clustering, and CMB lensing measurements from DES, \emph{Planck}, and SPT. Although the DES SNe results have been challenged based on differences with the Pantheon+ compilation \citep{Efstathiou25}, these differences can be largely explained by well-motivated methodological choices \citep{Vincenzi25}. Galaxy lensing and clustering measurements have long been associated with a tension in $S_8$ \citep[e.g.][]{Abdalla22,DiValentino25}, but some recent results (including the ones used in this paper) show consistency with the CMB \citep{Abbott23,Chen24,DESKiDS23,DES24b,DESI24.KP7B,Wright25}. Future constraints on $S_8$ and $\Omega_\mathrm{m}$ from DESI, \emph{Euclid}, and LSST could adjudicate this issue.
\end{itemize}

On the other hand, with an optical depth as low as $\tau=0.0491^{+0.0014}_{-0.0031}$, the constraints from DESI BAO and the CMB are in tension. This gives rise to several anomalies or apparent departures from $\Lambda$CDM. We have shown that the Lyman-$\alpha$ constraint on $\tau$ strengthens the evidence for the gravitational lensing excess, unphysical neutrino masses, and dynamical dark energy with an evolving equation of state. It remains an open questions whether these hints truly point to a new cosmological paradigm \citep{Ishak25,Leauthaud25,Efstathiou25b}. Future CMB experiments could significantly reduce the uncertainty in $\tau$ \citep{LiteBIRD23}. Analyses of the 21-cm signal, the patchy kSZ effect, and the Lyman-$\alpha$ forest provide other promising avenues. Ongoing observations probing the early stages of reionization will bolster the robustness of astrophysical constraints on $\tau$, which may prove pivotal in testing our currently favoured cosmological model. 


\section*{Acknowledgements}

I am grateful for useful and insightful comments by the anonymous referee. I also thank Mengtao Tang and Richard Ellis for providing the posterior distributions from their analysis of the redshift-dependent neutral fraction. I thank Carlos Frenk, Adrian Jenkins, Baojiu Li, and Silvia Pascoli for encouragement and useful discussions. I acknowledge support from the Royal Astronomical Society and STFC Consolidated Grant ST/X001075/1. This work used the DiRAC@Durham facility managed by the Institute for Computational Cosmology on behalf of the STFC DiRAC HPC Facility (www.dirac.ac.uk). The equipment was funded by BEIS capital funding via STFC capital grants ST/K00042X/1, ST/P002293/1 and ST/R002371/1, Durham University and STFC operations grant ST/R000832/1. DiRAC is part of the National e-Infrastructure.

\section*{Data Availability}

We publicly release our likelihood code at \url{https://github.com/wullm/reionlik}. The data will be made available upon request.



\bibliographystyle{mnras}
\bibliography{main} 




\appendix

\section{Technical details}\label{sec:details}

To map the Gaussian process to the interval $(0,1)$, we use a wrapping function $F\colon\mathbb{R}\to(0,1)$. For this purpose, we choose the cumulative distribution function of the standard generalized normal distribution with shape parameter $\beta=8$,
\begin{align}
    F(x) = \frac{1}{2} + \text{sign}(2x-1)\frac{1}{2\Gamma(1/\beta)}\gamma\left(1/\beta,\left\rvert2x-1\right\rvert^\beta\right),
\end{align}

\noindent
where $\gamma(s,x)=\int_0^xt^{s-1}e^{-t}\mathrm{d}t$ is the unnormalized incomplete lower gamma function. This function is very close to linear over most of its range, ensuring that $\chi(z)\approx x_\mathrm{HI}(z)$.

Inspired by the Gaussian process reconstruction of the dark energy equation of state, $w(z)$, of \citet{Lodha25}, we impose a generalized inverse Gaussian prior on the quantity $100\ell$,
\begin{align}
    f(x\;\rvert\; p,b) = \frac{b^{p/2}}{2K_p(\sqrt{b})}x^{p-1}e^{-(x+b/x)/2}, \qquad x>0,
\end{align}

\noindent
where $p=3$ and $b=1.5$ and where $K_p$ is the modified Bessel function of the second kind. This choice penalizes functions with too much freedom at low values of $\ell$ without oversampling nearly linear functions with high values of $\ell$.

Similar to \citet{Lodha25}, we introduce a latent variable corresponding to the value of $x_\mathrm{HI}(z_\mathrm{lat})$ at $z_\mathrm{lat}=7$. We then generate the Gaussian process conditional on this variable, which accelerates convergence. We perform our Gaussian process regression using a modified version of the \texttt{CLASS} code \citep{Lesgourgues11} that allows for a dense sampling of the free electron fraction, $x_\mathrm{e}(z)$, over the redshift range $[z_\mathrm{min},z_\mathrm{max}]$.

\section{Alternative dataset}\label{sec:tang_fit}

\begin{figure}
    \includegraphics[width=.48\textwidth]{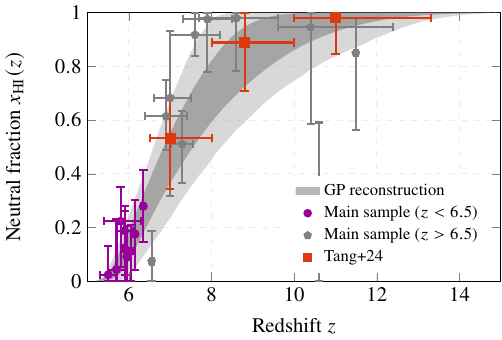}
    \caption{Gaussian process reconstruction of the reionization history between $5<z<15$, using an alternative set of constraints, consisting of the low-redshift data points from the main sample for $z<6.5$ and the high-redshift data points from \citet{Tang24} for $z>6.5$. The dark and light regions represent the 68\% and 95\% posterior predictive distributions.
    }
    \label{fig:tang_reconstruction}
\end{figure}

We present an alternative Gaussian process reconstruction of the reionization history based on the Lyman-$\alpha$ equivalent width constraints from \citet{Tang24}. After replacing all data points at $z>6.5$ from our main sample with the three constraints from \citet{Tang24}, we obtain the reconstructed histories shown in Fig.~\ref{fig:tang_reconstruction}. The analysis still favours a late reionization history, with a midpoint at $z_\mathrm{mid}=7.21^{+0.24}_{-0.44}$ and a duration of $\Delta z_{50}=1.97^{+0.46}_{-0.64}$. The duration is more extended compared to the main analysis, which yields $\Delta z_{50}=1.09^{+0.10}_{-0.28}$, albeit with significantly greater uncertainty due to the reduction in the number of data points. Overall, we conclude that the results are consistent with the main analysis.

\label{lastpage}
\end{document}